\documentclass[12pt]{article}

\usepackage{graphicx}
\usepackage{subfigure,epsfig,epstopdf}
\usepackage{amsfonts,amsmath,amssymb,amsthm}
\usepackage{indentfirst}
\usepackage{cite}
\usepackage{hyperref}
\usepackage{enumitem}
\usepackage{mathptmx}

\numberwithin{equation}{section}
\newcommand{\email}[1]{\footnote{E-mail: \href{mailto:#1}{#1}}}
\begin{document}

\title{SIM$(1)$--VSR Maxwell-Chern-Simons electrodynamics}
\author{R. Bufalo$^{1}$\email{rbufalo@ift.unesp.br} \\
%EndAName
\textit{$^{1}$ \small Instituto de F\'{\i}sica Te\'orica (IFT), Universidade Estadual Paulista} \\
\textit{\small Rua Dr. Bento Teobaldo Ferraz 271, Bloco II, 01140-070 S\~ao Paulo, SP, Brazil}\\
}
\maketitle
\date{}

\begin{abstract}
In this paper we propose a very special relativity (VSR)-inspired generalization of the Maxwell-Chern-Simons (MCS) electrodynamics.
This proposal is based upon the construction of a proper study of the SIM$(1)$--VSR gauge-symmetry.
It is shown that the VSR nonlocal effects present a significant and health departure from the usual MCS theory.
The classical dynamics is analysed in full detail, by studying the solution
for the electric field and static energy for this configuration.
Afterwards, the interaction energy between opposite charges are derived and we show that the VSR effects
play an important part in obtaining a (novel) finite expression for the static potential.
\end{abstract}

\begin{flushleft}
Very special relativity; SIM$(1)$ gauge-symmetry; Maxwell-Chern-Simons theory; Classical solutions
\end{flushleft}

%%%%%%%%%%%%%%%%%%%%%%%%%%%%%%%%%%%%%%%%%%%%%%%%%%%%%%%%%%%%%%%%%%%%%%%%%%%%%%%%%%%%%%%%%%%%%%%%%%%%%%%%%%%%%%%%%%%%%%%%%%%%%%%%%%%%%%%%%%%%%%
\newpage

\tableofcontents

%%%%%%%%%%%%%%%%%%%%%%%%%%%%%%%%%%%%%%%%%%%%%%%%%%%%%%%%%%%%%%%%%%%%%%%%%%%%%%%%%%%%%%%%%%%%%%%%%%%%%%%%%%%%%%%%%%%%%%%%%%%%%%%%%%%%%%%%%%%%%%%%%%%%%%%%
\section{Introduction}
\label{sec1}

In recent years we have been scrutinizing Planck scale Physics through many theories, proposals, ideas, etc, all this effort expended
in order to improve our understanding of the Nature behaviour at shortest distances (as well as in the beginning our Universe \cite{ref33}).
In particular, it is widely aimed by these proposals achieve a better description of a quantum theory of gravity,
or at least to make contact with the phenomenology of quantum gravity and hence gain insights about the fundamental structure of space and time at Planck scale.

Within the class of theories to describe Planck scale Physics, at the quantum realm,
we can cite String theory and loop quantum gravity as those most prominent candidates
up-to-date. Our main interest is to put at the same level a steady description of both quantum mechanics
and general relativity. An interesting outcome of these proposals is the presence of a minimal
measurable length scale \cite{ref22}, this can be incorporate at the quantum theory by the so-called
Generalized Uncertainty Principle \cite{ref20,ref21,ref23}.
Another consequence of known theories of quantum gravity is the breaking of (some) symmetry groups.
In particular, a well known scrutinized consequence is the violation of the underlying Lorentz
symmetry \cite{ref37,ref53}, since a definitive description of the space-time is expected not to be in terms of a classical smooth geometry.

Among the broad class of attempts trying to encompass and describe consistently Lorentz
violating effects \cite{ref25,ref26,ref27,ref28}, we shall focus in exploring features
of very special relativity (VSR) \cite{ref7,ref30} in this paper.
The cornerstone from this proposal is that the laws of physics are not invariant under the whole Poincar\'e group
but rather are invariant under subgroups of the Poincar\'e group preserving the basic elements of
special relativity, but at the same time enhancing the Lorentz algebra by modifying the dynamics of particles.
For instance, conservation laws and the usual relativistic dispersion relation,
$E^2=p^2+M^2$ for a particle of mass M, etc, are preserved in this case.

In particular, within this proposal, one can use in the realization of VSR the representations of the full Lorentz
group but supplemented by a Lorentz-violating factor, such that the symmetry of the Lagrangian is then reduced to one of the VSR
subgroups of the Lorentz group. 
These effects can then be encoded in the form Lorentz-violating terms in the Lagrangian that are
necessarily nonlocal.
As an example, one can observe that a VSR-covariant Dirac equation has the form 
\begin{equation}
\left(i\gamma^\mu \tilde{\partial}_\mu -m \right)\Psi \left(x\right)=0,
\end{equation}
where the wiggle operator is defined such as $\tilde{\partial}_{\mu}=\partial_{\mu}+\frac{1}{2}\frac{\sigma^{2}}{n.\partial}n_{\mu}$,
with the chosen preferred null direction $n_{\mu}=\left(1,0,0,1\right)$ so that it transforms multiplicatively
under a VSR transformation. Next, we can square the VSR-covariant Dirac equation, and we find
\begin{equation}
\left(\partial^\mu \partial_\mu +\mathcal{M}^2 \right)\Psi \left(x\right)=0, \quad \mathcal{M}^2=M^2+\sigma^2 .
\end{equation}
We thus immediately realize an interesting observable consequence of VSR that is to provide a novel mechanism
for introducing neutrino masses without the need for new particles \cite{ref30}.
Moreover, the VSR parameter $\sigma$ sets the scale for the VSR effects.
Among the most interesting analysis involving VSR effects we can cite a realization of VSR via
a lightlike noncommutative deformation of Poincar\'e symmetry \cite{ref6}, studies on Dirac equation \cite{ref13}
and hydrogen atom \cite{ref3}, as well as gauge theories \cite{ref16} and curved spacetime field theories 
\cite{ref9}, gravitational and cosmological models \cite{ref8,ref36}. 

As it concerns our interest, VSR-effects have been discussed in the context of $(3+1)$-dims electromagnetic theories:
Abelian and non-Abelian Maxwell theories \cite{ref4,ref5,ref51}, Chern-Simons theory \cite{ref2,ref52} and Born-Infeld electrodynamics \cite{ref1}.
By different reasons, we have seen recently a renewed interest in studying Lorentz-violating modifications of the Maxwell-Chern-Simons theory \cite{ref2,ref52,ref32,ref12}.
Although of the interesting features obtained in those analysis, one may wonder what the (nonlocal) VSR-effects
may influence the behaviour of a lower-dimensional electromagnetic theory, for instance in a $(2+1)$-dimensional spacetime \cite{ref15},
where we will work with the SIM$(1)$ subgroup (of the $SO(2,1)$ Lorentz group) that preserve all the aforementioned conditions,
in particular that a given null-vector is preserved up to a rescaling. \footnote{A detailed account of the SIM$(1)$ subgroup can be found in Ref.~\cite{ref51}}

It worth notice that different approaches have been used to consider mass effects in
$(2+1)$-dims generalized electrodynamic theories \cite{ref45,ref46}.
It is well-known that the Maxwell-Chern-Simons (MCS) electrodynamics describe a single massive gauge mode
of helicity $\pm1$, the so-called Topologically massive electrodynamics \cite{ref42}.
Hence, we expect that the VSR setting will not modify Maxwell-Chern-Simons theory only by adding 'massive' effects, \footnote{Notice
however that although VSR engender a nonzero mass, it preserves the number of polarization states \cite{ref4}.}
since a topological mass is already present, but rather that the VSR-effects will be prominent in changing the
theory's dynamics in a significant and novel manner.

In this letter we will examine the Maxwell-Chern-Simons electrodynamics in a VSR setting.
We start Sec.\ref{sec2} by reviewing the fundamental aspects from the SIM$(1)$--VSR gauge invariance,
which allow us to determine the VSR-modified Abelian field strength to be used in our analysis.
In Sec.\ref{sec3}, we define our SIM$(1)$--VSR Maxwell-Chern-Simons theory. Afterwards, we determine the dispersion relation
and discuss the electrostatic solution for the equations of motion in the presence of a pointlike charge.
In addition, we compute the field energy and gauge-invariant potential between two opposite charges.
Along the analysis we will comment at pertinent points the differences obtained by VSR deformations
in view of the usual MCS theory.
In Sec.\ref{sec4} we summarize the results, and present our final remarks.

%%%%%%%%%%%%%%%%%%%%%%%%%%%%%%%%%%%%%%%%%%%%%%%%%%%%%%%%%%%%%%%%%%%%%%%%%%%%%%%%%%%%%%%%%%%%%%%%
%%%%%%%%%%%%%%%%%%%%%%%%%%%%%%%%%%%%%%%%%%%%%%%%%%%%%%%%%%%%%%%%%%%%%%%%%%%%%%%%%%%%%%%%%%%%%%%%
\section{SIM$(1)$ gauge symmetry overview}
\label{sec2}

Let us start by discussing the SIM$(1)$ VSR gauge invariance \cite{ref4,ref5}.
An important remark to bear is that although the VSR subgroups do not admit invariant tensors, they select a preferred null direction.
For this matter, the first point in order to develop the gauge invariance is to realize that the gauge transformation
of a gauge field in VSR is modified so that
\begin{equation}
\delta A_{\mu}=\tilde{\partial}_{\mu}\Lambda,\label{eq:1.1}
\end{equation}
where the wiggle operator is defined such as $\tilde{\partial}_{\mu}=\partial_{\mu}+\frac{1}{2}\frac{\sigma^{2}}{n.\partial}n_{\mu}$,
but now with the chosen preferred null direction given as $n_{\mu}=\left(1,0,1\right)$ and multiplicatively
covariant under the SIM$(1)$ subgroup of the $(2+1)$-dims Lorentz group \cite{ref51}.

Next, let us consider a charged scalar field $\varphi$ with an infinitesimal
gauge transformation given as usual by $\delta\varphi=i\Lambda\varphi$. Moreover,
we know that in general a covariant derivative must satisfy the transformation property
\begin{equation}
\delta\left(D_{\mu}\varphi\right)=i\Lambda\left(D_{\mu}\varphi\right).\label{eq:1.2}
\end{equation}
Hence, it can be showed that the covariant operator defined as the following
\begin{equation}
D_{\mu}\varphi=\partial_{\mu}\varphi-iA_{\mu}\varphi+\frac{i\sigma^{2}}{2}n_{\mu}\left(\frac{1}{\left(n.\partial\right)^{2}}\left(n.A\right)\right)\varphi\label{eq:1.3}
\end{equation}
satisfies the condition \eqref{eq:1.2}. Besides, in the same way as we have defined the wiggle operator $\tilde{\partial}$
from the raw derivative $\partial$, we can generalize the covariant derivative $D$ to a wiggle operator 
\begin{equation}
\tilde{D}_{\mu}=D_{\mu}+\frac{1}{2}\frac{\sigma^{2}}{n.D}n_{\mu}\label{eq:1.4}
\end{equation}
so that it reduces to the operator $\tilde{\partial}$ when the limit $A_{\mu}=0$ is taken. 

Hence, with the above definitions the field strength associated to the operator $D_{\mu}$ can be computed as usual
by the following quantity $\left[D_{\mu},D_{\nu}\right]\varphi=-iF_{\mu\nu}\varphi$.
This can be shown to result into
\begin{equation}
F_{\mu\nu}=\partial_{\mu}A_{\nu}+\frac{\sigma^{2}}{2}n_{\mu}\left(\frac{1}{\left(n.\partial\right)^{2}}\partial_{\nu}\left(n.A\right)\right)-\mu\leftrightarrow\nu\label{eq:1.5}
\end{equation}
This field strength can be seen as the raw definition of the $A_{\mu}$ gauge field strength.
However, one can easily realize that this field-strength does not coincide with the SIM$(1)$ wiggle operator
\begin{equation}
\tilde{F}_{\mu\nu}=\tilde{\partial}_{\mu}A_{\nu}-\tilde{\partial}_{\nu}A_{\mu}\label{eq:1.6}
\end{equation}
On one hand, the wiggle definition $\tilde{F}_{\mu\nu}$ is gauge invariant and it will be used
to describe massive gauge fields. Now, on the other hand, we can realize
that the difference between the raw and wiggle field-strength must
be gauge invariant as well, so that we can write the following expression
for wiggle field strength
\begin{equation}
\tilde{F}_{\mu\nu}=F_{\mu\nu}+\frac{\sigma^{2}}{2}\frac{1}{\left(n.\partial\right)^{2}}\left(n_{\nu}n^{\lambda}F_{\mu\lambda}
-n_{\mu}n^{\lambda}F_{\nu\lambda}\right)\label{eq:1.7}
\end{equation}

Some remarks are now in place. By means of illustration, in showing how to describe massive gauge
fields, let us consider a VSR modified Maxwell action,
\begin{equation}
S=\int d^{\omega}x\left[-\frac{1}{4}\tilde{F}_{\mu\nu}\tilde{F}^{\mu\nu}\right]\label{eq:1.8}
\end{equation}
it is interesting to notice that this action can be augmented by further quadratic terms in \textbf{$A$}
as well as by gauge invariant coupling to matter fields \cite{ref5}.
In particular, this prescription also works to generates mass for the matter fields.
The field equations follow straightforwardly as
\begin{equation}
\tilde{\partial}_{\mu}\tilde{F}^{\mu\nu}=0,
\end{equation}
now, by taking a VSR-type Lorenz gauge condition, $\tilde{\partial}_{\mu}A^{\mu}=0$, we find that
\begin{equation}
\tilde{\partial}^{2}A^{\mu}=\left(\partial^{2}+m^{2}\right)A^{\mu}=0.
\end{equation}

With this discussion we see that a massive gauge field, defined in terms of the ordinary derivative, can be described
suitably in a gauge-invariant fashion when written in terms of the wiggle operator.
Moreover, this may be considered our starting point in defining our model of interest.

%%%%%%%%%%%%%%%%%%%%%%%%%%%%%%%%%%%%%%%%%%%%%%%%%%%%%%%%%%%%%%%%%%%%%%%%%%%%%%%%%%%%%%%%%%%%%%%%%%%%%%%%%%%%%%%%%%%%%%%%%%%%%%%%%%%%%%%%%%%%%%%%%%%%%%%%%%%%%%%%%%%%%%%%%%%%%%%%%%%%%%%%%%%%%%%%%%%%%%%%%%%%%%%%%%%%%%%%%%%%%%%%%%%%%%%%%%%%%%%%%%%%%%%%%%%%%%%%%%%%%%%%
\section{VSR Maxwell-Chern-Simons electrodynamics}
\label{sec3}

Let us now characterize the model under consideration. Based on the points discussed above,
but taking into account a SIM$(1)$ VSR setting and the wiggle field strength expression \eqref{eq:1.7}, we are in a
position to define the SIM$(1)$--VSR Maxwell-Chern-Simons electrodynamics by the following Lagrangian density
\begin{equation}
\mathcal{L}=-\frac{1}{4}\tilde{F}_{\mu\nu}\tilde{F}^{\mu\nu}+\frac{m}{4}\varepsilon^{\mu\nu\lambda}A_{\mu}\tilde{F}_{\nu\lambda}.\label{eq:2.1}
\end{equation}
The usual MCS theory, or topologically massive electrodynamics, describe a single massive gauge mode of helicity $\pm1$.
We shall now explore the VSR setting in order to look for modification on the solutions of the MCS classical dynamics.
Next, the equations of motion for the SIM$(1)$ MCS theory can be readily determined as
\begin{equation}
\tilde{\partial}_{\mu}\tilde{F}^{\mu\alpha}+\frac{m}{2}\varepsilon^{\alpha\nu\lambda}\tilde{F}_{\nu\lambda}=0\label{eq:2.2}
\end{equation}
In order to solve the above equations, it is convenient to introduce the dual field strength $\tilde{G}_{\mu}$,
which is a vector in three dimensions $\tilde{G}^{\mu}=\frac{1}{2}\varepsilon^{\mu\nu\lambda}\tilde{F}_{\nu\lambda}$.
Moreover, it follows straightforwardly that the Bianchi identity in this case is written as $\partial_{\mu}\tilde{G}^{\mu}=0$.
Hence, we see that the field equations \eqref{eq:2.2} are now written in the form
\begin{equation}
\left[\varepsilon^{\mu\nu\lambda}\tilde{\partial}_{\mu}+m\eta^{\nu\lambda}\right]\tilde{G}_{\lambda}=0\label{eq:2.3}
\end{equation}
From this expression we can identify the (on-shell) projection operators \cite{ref42}
\begin{equation}
\left[P\left(\pm m\right)\right]_{\nu}^{\mu}=\frac{1}{2}\left[\delta_{\nu}^{\mu}\mp\frac{1}{m}\varepsilon_{ ~~~~\nu}^{\mu\lambda}\tilde{\partial}_{\lambda}\right],
\end{equation}
it is easy to show that, as expected, they satisfy $\left[P\left(\pm m\right)\right]^{2}=\left[P\left(\pm m\right)\right]$.
Actually, these operators project onto the Poincar\'e (irreducible) representations \cite{ref42}.
Hence, in terms of the dual field strength $\tilde{G}_{\mu}$, it finally follows the field equation
\begin{equation}
\left(\partial^{2}+M^{2}\right)\tilde{G}_{\mu}=0\label{eq:2.4}
\end{equation}
where we have defined a new mass parameter $M^{2}=m^{2}+\sigma^{2}$.
This shows, nonetheless, that the dispersion relation for the gauge field is only slightly modified,
since the dispersion relation $\omega=\pm\sqrt{p^{2}+M^{2}}$ has the same form as the one obtained in the usual theory,
being only shifted on the mass parameter.

By means of discussion, let us now add a (electrostatic) source term
$A_{0}J^{0}$ into the Lagrangian \eqref{eq:2.1}. Thus, a new set of field equations
now read
\begin{equation}
\tilde{\partial}_{\mu}\tilde{F}^{\mu\alpha}+\frac{m}{2}\varepsilon^{\alpha\nu\lambda}\tilde{F}_{\nu\lambda}=J^{0}\delta_{0}^{\alpha}. \label{eq:2.5}
\end{equation}
Hence, for the temporal component of \eqref{eq:2.5}, we find a modified Gauss's law
\begin{equation}
\tilde{\partial}_{i}\tilde{E}^{i}+\frac{m}{2}\tilde{B}=J^{0}\label{eq:2.6}
\end{equation}
where we have defined the wiggle electric and magnetic fields such as $\tilde{E}^{i}=\tilde{F}^{i0}$
and $\tilde{B}=\frac{1}{2}\varepsilon^{ij}\tilde{F}_{ij}$, respectively. Besides, it follows
that for $\alpha=i$ in \eqref{eq:2.5}, we have the relation 
\begin{equation}
\tilde{E}_{i}=\frac{1}{m}\tilde{\partial}_{i}\tilde{B} .\label{eq:2.7}
\end{equation}
Finally, we can use the relation \eqref{eq:2.7} to rewrite \eqref{eq:2.6} in the following form,
\begin{equation}
\left(-\nabla^{2}+M^{2}\right)\tilde{B}=mJ^{0}. \label{eq:2.8}
\end{equation}

In particular, we can consider a simple scenario in order to solve \eqref{eq:2.8} (i.e. Eq.\eqref{eq:2.6}), this can be chosen
by taking the current density for a pointlike charge $J_{0}\left(t,\mathbf{r}\right)=g\delta^{\left(3\right)}\left(\mathbf{r}\right)$.
Hence, one can easily solve \eqref{eq:2.8} to find
\begin{align}
\tilde{B}\left(\mathbf{r}\right) & =\frac{gm}{2\pi}K_{0}\left(Mr\right).\label{eq:2.9}
\end{align}
Finally, we can determine the electric field by replacing \eqref{eq:2.9} back into \eqref{eq:2.7}
\begin{equation}
\tilde{{\bf E}}\left(\mathbf{r}\right) =-\frac{gM}{2\pi}K_{1}\left(Mr\right)\left(\tilde{\nabla} r\right). \label{eq:2.15}
\end{equation}
One can see that the wiggle derivative results into $\tilde{\nabla}r = \hat{{\bf r}}
-\frac{\sigma^{2} \hat{{\bf n}}}{2}\left(\frac{1}{\nabla_{y}}r\right)$, where the unit vector is given as $\hat{{\bf n}}=\left(0,1\right)$.
Let us now concentrate in computing the nonlocal term of the above expression. This can be worked out by means of the following representation
\begin{equation}
\frac{\sigma}{\nabla_{y}}r=\int_{0}^{\infty}ds\left(\sum_{n=0}\frac{1}{n!}\left(-\frac{s\nabla_{y}}{\sigma}\right)^{n}\right)\sqrt{x^{2}+y^{2}}=\int_{0}^{\infty}ds\sqrt{x^{2}+\left(\frac{s}{\sigma}-y\right)^{2}} . \label{eq:2.16}
\end{equation}
Besides, the above derivative has been calculated by means of standard manipulations:
One can make use of Newton's binominal to rewrite $\sqrt{x^{2}+y^{2}}$ conveniently as $\sum_{k}\left(\begin{array}{c}
\frac{1}{2} \\ k \end{array} \right)\left(x^{2}\right)^{1/2-k}\left(y^{2}\right)^{k} $, so that we can compute the operation
$\frac{1}{n!}\frac{d^{n}}{dy^{n}}\left(y^{2}\right)^{k}=\left(\begin{array}{c}2k\\ n \end{array}\right)y^{2k-n}$.
Finally, one can solve the integration in \eqref{eq:2.16} and find that
\begin{align}
\frac{\sigma}{\nabla_{y}}r &  =\frac{\sigma}{2}\left[yr-x^{2}\ln\left(\sigma\left[r-y\right]\right)\right]+\lim _{\rho \rightarrow \infty}\Lambda\left(\rho\right). \label{eq:2.17}
\end{align}
We thus see in \eqref{eq:2.17} that as a consequence of the nonlocality of the VSR-effects 
the distribution $\Lambda\left(\rho\right)\equiv \frac{1}{2}\left[-y \sqrt{\sigma^2x^{2}+\left(\sigma y-\rho\right)^{2}}+\sigma x^{2}\ln\left(\rho-\sigma y+\sqrt{\sigma^2x^{2}+\left(\sigma y-\rho\right)^{2}}\right)\right]$ is not regular, diverging as $\rho\rightarrow\infty$.
Nonetheless, in the first term of \eqref{eq:2.17}, we have a finite and well-behaved  contribution,
which we shall consider in our following analysis while disregarding the non-regular $\Lambda\left(\rho\right)$ contribution.
This approximation is valid since the finite part is sufficient to propagate VSR deviations.

Therefore, from the above discussion, we find that the wiggle electric field is then given by
\begin{equation}
\left|\tilde{\mathbf{E}}\right|=-\frac{gM}{2\pi}K_{1}\left(Mr\right)\left[1-\frac{\sigma^{2}\left(\hat{\mathbf{r}}.\hat{\mathbf{n}}\right)}{4}\left[yr-x^{2}\ln\left(\sigma\left[r-y\right]\right)\right]\right].\label{eq:2.12}
\end{equation}
The complete expression for the electric field \eqref{eq:2.12} can be rewritten in polar coordinates,
so that $\left(\hat{\mathbf{r}}.\hat{\mathbf{n}}\right)=\sin\theta$. Thus, we find that it now reads
\begin{equation}
\left|\tilde{\mathbf{E}}\right|=-\frac{gM}{2\pi}K_{1}\left(Mr\right)\left[1-\frac{\sigma^{2}r^{2}\sin\theta}{4}\left(\sin\theta-\cos^{2}\theta\ln\left[\sigma r\left(1-\sin\theta\right)\right]\right)\right]. \label{eq:2.10}
\end{equation}

By means of illustration, let us consider a fixed angle $\theta=\pi/2$, so that we can examine
the electric field short distance behaviour. With these considerations, we find 
\begin{equation}
\left|\tilde{\mathbf{E}}\right|=-\frac{gM}{2\pi}K_{1}\left(Mr\right)\left[1-\frac{\sigma^{2}r^{2}}{4}\right]\simeq -\frac{g}{2\pi}\left[\frac{1}{r}-\frac{\sigma^{2}}{4}r\right]. \label{eq:2.18}
\end{equation}
Hence, we see that the electric field $\left|\tilde{\mathbf{E}}\right|$ at the SIM$(1)$ MCS theory is
still non-regular at the origin, as $r\rightarrow0$, due to the usual MCS part.
Nonetheless, it is worth of mention that the SIM$(1)$--VSR contribution already gives a well-behaved and regular contribution.
We will observe further this positive consequence of VSR acting as a regulator of singular points when computing the interparticle potential (see Sec.\ref{sec:3.2}).

%%%%%%%%%%%%%%%%%%%%%%%%%%%%%%%%%%%%%%%%%%%%%%%%%%%%%%%%%%%%%%%%%%%%%%%%%%%%%%%%%%%%%%%%%%%%%%%%%%%%%%%%%%%%%%%%%%%%%%%%%%%%%%%%
%%%%%%%%%%%%%%%%%%%%%%%%%%%%%%%%%%%%%%%%%%%%%%%%%%%%%%%%%%%%%%%%%%%%%%%%%%%%%%%%%%%%%%%%%%%%%%%%%%%%%%%%%%%%%%%%%%%%%%%%%%%%%%%%
\subsection{Electrostatic energy}
\label{sec:3.1}

Is of our interest to proceed and compute the total amount of energy stored in the
electrostatic field of a pointlike charge, $U=\int d^{2}xT_{0}^{0}$.
The energy-momentum tensor can be evaluated as usual $T_{\mu\nu}=\frac{2}{\sqrt{-g}}\frac{\delta\left(\sqrt{-g}\mathcal{L}\right)}{\delta g^{\mu\nu}}$.
However, notice that the Chern-Simons contribution, $\int dx\varepsilon^{\mu\nu\lambda}A_{\mu}\tilde{F}_{\nu\lambda}$,
is already coordinate invariant \cite{ref42}, without additional metric factors;
so that the CS mass term does not contribute to $T_{\mu\nu}$ (as expected from a topological term).
Hence, we find in our case, that the energy-momentum tensor is simply given as
\begin{equation}
T_{\nu}^{\mu}=-\tilde{F}^{\mu\lambda}\tilde{F}_{\nu\lambda}+\frac{\delta_{\nu}^{\mu}}{4}\tilde{F}_{\sigma\lambda}\tilde{F}^{\sigma\lambda}.
\end{equation}
So, in the electrostatic limit we find $T_{0}^{0}=\frac{1}{2}\left|\tilde{\mathbf{E}}\right|^{2}$.
Thus, by using the solution \eqref{eq:2.10} we have that
\begin{equation}
U =\frac{g^{2}M^{2}}{8\pi^{2}}\int rdr\left(K_{1}\left(Mr\right)\right)^{2}\int_{0}^{2\pi}d\theta\left[1-\frac{\sigma^{2}r^{2}\sin\theta}{4}\left[\sin\theta-\cos^{2}\theta\ln\left[\sigma r\left(1-\sin\theta\right)\right]\right]\right]^{2} .
\end{equation}
The angular integration can be computed by means of standard results, so that we get
\begin{align}
U & =\frac{g^{2}M^{2}}{8\pi}\int r^{1-\varepsilon}dr\left(K_{1}\left(Mr\right)\right)^{2}\biggl[2-\frac{2}{3}\sigma^{2}r^{2}\nonumber \\
&+\frac{\sigma^{4}r^{4}}{9216}\biggl( 469-60\ln2 +72\left(\ln \sigma r\right)^{2}+12\left(5-12\ln2\right)\ln \sigma r\biggr)\biggr]
\end{align}
where we have introduced into the numerator a $r^{-\varepsilon}$ factor, as $\varepsilon \rightarrow 0$,
so that we can compute the radial integration exactly.
A straightforward computation of the remaining integration results into the following expression for the field energy
\begin{align}
U &=\frac{g^{2}}{8\pi}\biggl[-\frac{2}{\varepsilon}-\left(1+2\gamma+2\ln\frac{M}{2}\right) -\frac{4\sigma^{2}}{9M^{2}}\nonumber \\
&+\frac{\sigma^{4}}{72000M^{4}}\biggl(6922+90\gamma\left(-21+10\gamma\right)-900\left(\ln2\right)^{2}+90\ln \left( \frac{M}{\sigma}\right)\left(-21+20\gamma+10\ln \left( \frac{M}{\sigma}\right)\right)\biggr)\biggr].\label{eq:2.11}
\end{align}

We thus find a regularized divergence in the first term of the field energy \eqref{eq:2.11}; moreover, we clearly see that this divergent term is inherent from the usual MCS theory
(a similar fact is also present in the Maxwell theory at $(3+1)$-dims.).
So, in order to compare the field energy between the MCS and SIM$(1)$--VSR MCS theories,
we shall consider only the finite contribution from the energy expression \eqref{eq:2.11}.
First, for the VSR parameter $\sigma=0$, we find the usual MCS (finite) contribution
\begin{align}
U^{MCS} & =-\frac{g^{2}}{8\pi}\biggl[1+2\gamma+2\ln\frac{m}{2}\biggr]
\end{align}
while, the SIM$(1)$--VSR (finite) contribution $U^{VSR}$ follows by taking $m=0$ in \eqref{eq:2.11}, i.e. $M=\sigma$.

\begin{figure*}[tbp]
\centering

{\epsfig{figure=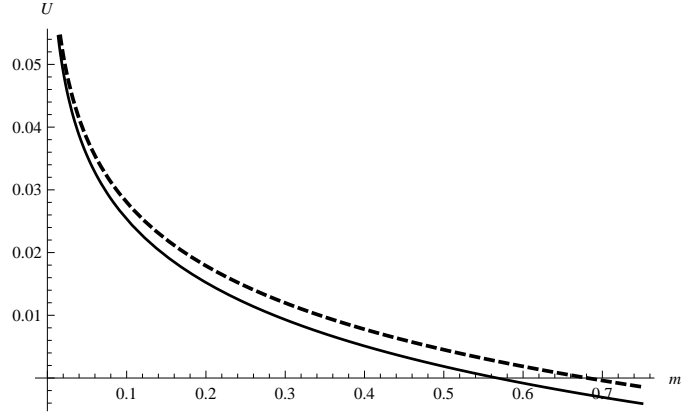,width=3.5in}} 

\caption{The solid and dashed lines correspond to the $U^{VSR}$ and $U^{MCS}$ contributions, respectively, written in terms of the mass.
Showing the mass range for both contributions for a positive energy. 
}
\label{fig1}
\end{figure*}

We can easily compute and find 
 $U^{MCS}$ has one zero point
in $m=0.681085$, while $U^{VSR}$ has one zero points in $m_{1}=0.567385$.
So the VSR-modified MCS contribution has a shorter range of positivity than the usual MCS contribution.
This is depicted in the Figure \ref{fig1}.

%%%%%%%%%%%%%%%%%%%%%%%%%%%%%%%%%%%%%%%%%%%%%%%%%%%%%%%%%%%%%%%%%%%%%%%%%%%%%%%%%%%%%%%%%%%%%%%%%%%%%%%%%%%%%%%%%%%%%%%%%%%%%%%%%%%%%%%%%%%%%%%%%%%%%%

%%%%%%%%%%%%%%%%%%%%%%%%%%%%%%%%%%%%%%%%%%%%%%%%%%%%%%%%%%%%%%%%%%%%%%%%%%%%%%%%%%%%%%%%%%%%%%%%%%%%%%%%%%%%%%%%%%%%%%%%%%%%%%%%%%%%%%%%%%%%%%%%%%%%%%
%%%%%%%%%%%%%%%%%%%%%%%%%%%%%%%%%%%%%%%%%%%%%%%%%%%%%%%%%%%%%%%%%%%%%%%%%%%%%%%%%%%%%%%%%%%%%%%%%%%%%%%%%%%%%%%%%%%%%%%%%%%%%%%%%%%%%%%%%%%%%%%%%%%%%%

\subsection{Static potential}
\label{sec:3.2}

In this last part of our analysis we will compute the VSR contribution for the static potential energy $V$ between pointlike sources.
This study is well motivated since it is usually chosen to describe bound states of particle-antiparticle pairs.
Moreover, we will show that the VSR-effects can be chosen conveniently so that the potential is well-behaved and regular.
A suitable framework to compute the potential is found to be in terms of physical gauge-invariant objects \cite{ref40,ref41}.
Let us start by defining the vector gauge-invariant field by
\begin{equation}
\mathcal{A}_{\mu}\left(x\right)=A_{\mu}\left(x\right)-\partial_{\mu}\int_{\mathcal{C}_{\xi x}}dz^{\lambda}A_{\lambda}\left(z\right),\label{eq:3.1}
\end{equation}
where the contour $\mathcal{C}_{\xi x}$ is chosen such as a spacelike path
from the point $\xi$ and $x$, on a fixed slice time.
Without loss of generality, we consider here a straight path $z_{i}=\xi_{i}+\zeta\left(x-\xi\right)_{i}$
parametrized by $\zeta$ ($0\leq\zeta\leq1$); besides, we can take by simplicity the fixed (reference) point to be $\xi_{i}=0$.
This construction for a gauge-invariant variable is, in fact, closely related to the Poincar\'e
gauge conditions $A_{0}\approx0$ and $\int_{\mathcal{C} }dz^{\lambda}A_{\lambda}\approx0$.

Within our interest, we can work out the expression \eqref{eq:3.1} under the above consideration, and after some manipulation, we find that its temporal component reads \cite{ref40}
\begin{equation}
\mathcal{A}_{0}\left(t,\mathbf{r}\right)=\int_{0}^{1}d\zeta x^{i}E_{i}\left(t,\zeta\mathbf{r}\right). \label{eq:3.2}
\end{equation}

A remark is now in place. On one hand, the interaction energy $V$ of a quantum mechanical system
is usually computed by means of a perturbative analysis, i.e. $\left\langle H\right\rangle _{\Omega}=\left\langle H\right\rangle _{0}+V$,
where the complete Hamiltonian is obtained by a canonical analysis following Dirac's procedure.
Moreover, in this case one have Dirac's gauge-invariant fermion--antifermion physical state
$\left|\Omega\right\rangle \equiv\left|\overline{\Psi}\left(\mathbf{0}\right)\Psi\left(\mathbf{L}\right)\right\rangle $.
On the other hand, instead, we may equally consider the gauge-invariant
field in \eqref{eq:3.2} as to provide an equivalent but rather simple
framework to compute the expression for the potential $V$ \cite{ref41}.

In particular, we can consider the scenario of a pair of static pointlike (opposite) charges,
i.e. $J^{0}\left(t,\mathbf{r}\right)=g\left[\delta^{\left(3\right)}\left(\mathbf{r}\right)-\delta^{\left(3\right)}\left(\mathbf{r}-\mathbf{L}\right)\right]$, where $L=\left|\vec{x}-\vec{y}\right|$. In this case, the potential is then defined as
\begin{equation}
V=g\left[\mathcal{A}_{0}\left(\mathbf{0}\right)-\mathcal{A}_{0}\left(\mathbf{L}\right)\right]\label{eq:3.7}
\end{equation}
Hence, in order to compute first the field $\mathcal{A}_{0}$ from \eqref{eq:3.2} we take the electric field solution Eq.\eqref{eq:2.12}.
After some straightforward manipulation, we get the following expression
\begin{align}
\mathcal{A}_{0}\left(t,\mathbf{r}\right)&=\frac{gMr}{2\pi}\int_{0}^{1}d\zeta K_{1}\left(\zeta Mr\right)\left[1-\zeta^{2}\frac{\sigma^{2}r^{2}}{4}\sin\theta\left[\sin\theta-\cos^{2}\theta\ln\left(\zeta \sigma r\left[1-\sin\theta\right]\right)\right]\right], \nonumber \\
& =\frac{g}{2\pi}\int_{0}^{Mr}dwK_{1}\left(w\right)\left[1+w^{2}\left[a^{2}\ln w-b^{2}\right]\right], \label{eq:3.10}
\end{align}
where we have made a change of variables $w=Mr\zeta$ and defined by simplicity
\begin{align}
a^{2} & =\frac{\sigma^{2}}{4M^{2}}\sin\theta\cos^{2}\theta ,\label{eq:3.11}\\
b^{2} & =\frac{\sigma^{2}}{4M^{2}}\sin\theta\left[\sin\theta+\cos^{2}\theta\ln\left(\frac{M}{\sigma}\frac{1}{1-\sin\theta}\right)\right]. \label{eq:3.12}
\end{align}
The integration in \eqref{eq:3.10} can be readily computed, and the complete expression for the gauge-invariant field reads
\begin{align}
\mathcal{A}_{0}\left(t,\mathbf{r}\right) & =-\frac{g}{2\pi}\biggl[\left(1+2a^{2}\right)\left.K_{0}\left(w\right)\right|_{0}^{Mr}+a^{2}\left(Mr\right)K_{1}\left(Mr\right)\nonumber \\
&+\left(Mr\right)\left(\ln\left(Mr\right)a^{2}-b^{2}\right)\left[\left(Mr\right)K_{0}\left(Mr\right)+2K_{1}\left(Mr\right)\right]\biggr]. \label{eq:3.3}
\end{align}
It is worth noticing the singular behaviour of $\left.K_{0}\left(w\right)\right|_{0}^{Mr}$ on \eqref{eq:3.3}.
Since the expansion of $K_{0}\left(w\right)$ for $w\rightarrow0$
goes as $K_{0}\left(w\right) \sim -\ln w$, we thus see that the lower limit from the first term is not regular.
This is indeed the case in the usual MCS theory, where such a term is usually disregarded.
Surprisingly, we see that the novel coefficient (VSR dependent) of this term can be chosen
conveniently in such a way that this divergence is removed.
Hence, for the case when the identity for coefficient holds $1+2a^{2}=0$, it yields for \eqref{eq:3.3}
\begin{align}
\mathcal{A}_{0}\left(t,\mathbf{r}\right)  =\frac{gMr}{4\pi}\biggl[K_{1}\left(Mr\right)+ \left(\ln\left(Mr\right)+2b^{2}\right)\left[\left(Mr\right)K_{0}\left(Mr\right)+2K_{1}\left(Mr\right)\right]\biggr]\label{eq:3.4}
\end{align}

\begin{figure*}[tbp]
\centering
\epsfig{figure=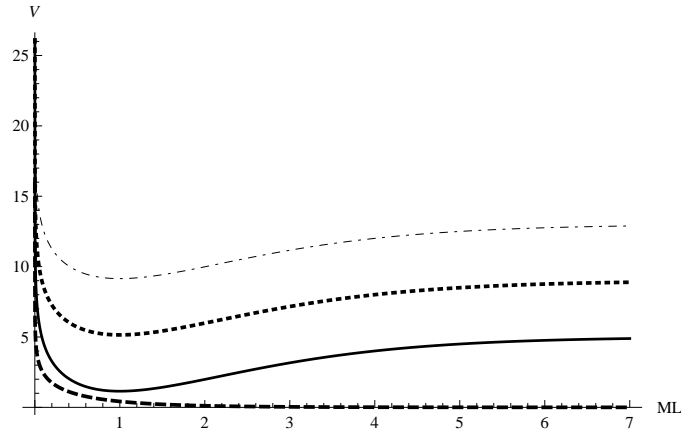,width=3.5in}

\caption{The dashed correspond to the $V^{MCS}$ contribution, while the solid, dotted, dotdashed lines correspond to the complete $V^{VSR}$ contributions, written in terms of $ML$
with an arbitrary choice of $b^2 =1$, $b^2 =2$ and $b^2 =3$, respectively.}
\label{fig2}
\end{figure*}

Otherwise, we can conceive this choice for the coefficient as if we are taking the following value for the VSR parameter
\begin{equation}
\sigma^{2}=-\frac{2m^{2}}{2+\sin\theta\cos^{2}\theta}, \label{eq:3.5}
\end{equation}
where we can think that such relation holds for a fixed value of $\theta = \sin^{-1} \left(\hat{\mathbf{r}}.\hat{\mathbf{n}}\right)$. In this case, it also follows that
\begin{align}
b^{2} =-\frac{1}{2}\left[\tan\theta\sec\theta+\ln\left(\frac{M}{\sigma}\frac{1}{1-\sin\theta}\right)\right].\label{eq:3.6}
\end{align}
Besides, the missing piece to evaluate the potential $V$ is obtained by taking the limit $r\rightarrow0$ in \eqref{eq:3.4},
$\mathcal{A}_{0}\left(\mathbf{0}\right) =\frac{g}{4\pi}\biggl[4b^{2}+1\bigg]$.
Finally, under the above considerations and by collecting the results and substituting them back into \eqref{eq:3.7}, we find for the potential the following result
\begin{align}
V^{VSR}&=-\frac{g^{2}}{4\pi}\bigg[ \left(ML\right)K_{1}\left(ML\right)-1-4b^{2}\nonumber \\
&+\left(ML\right)\left(\ln\left(ML\right)+2b^{2}\right)\left[\left(ML\right)K_{0}\left(ML\right)+2K_{1}\left(ML\right)\right]\bigg]. \label{eq:3.8}
\end{align}
At last, we see that the VSR deformed expression \eqref{eq:3.8} shows a significant departure from the usual behaviour of the MCS theory, see Figure \ref{fig2}.
By means of illustration, we can consider the short distance regime of the potential \eqref{eq:3.8},
i.e. $ML\ll 1$, this results into the simple (confining) expression
\begin{align}
V^{VSR} =-\frac{g^{2}}{2\pi}\biggl[\ln\left(M\left|\vec{x}-\vec{y}\right|\right)+\mathcal{O}\left(M^{2}L^{2}\right)\biggr].\label{3.9}
\end{align}

At first sight, this simplified expression might looks exactly the same as the one obtained in the usual MCS theory, since if we consider
the short distance regime we have $K_{0}\left(w\right) \sim -\ln w$ (see \eqref{eq:3.3} for $\sigma =0$).
However, notice two major differences: one, the VSR-modified potential \eqref{3.9} is completely regular and finite
under the condition $1+2a^{2}=0$, i.e., we have removed the term $K_{0}\left(0\right)$; second,
the effective mass $M^2=\sigma^2+m^2$ is shifted from the usual MCS parameter $m$.
At last, since the VSR deformed potential \eqref{eq:3.8} displays a confining behaviour at short distance (i.e. $V^{VSR} \rightarrow \infty$ as $L \rightarrow 0$), it can used to describe stable bound states of particle-antiparticle pairs.

%%%%%%%%%%%%%%%%%%%%%%%%%%%%%%%%%%%%%%%%%%%%%%%%%%%%%%%%%%%%%%%%%%%%%%%%%%%%%%%%%%%%%%%%%%%%%%%%%%%%%%%%%%%%%%%%%%%%%%%%%%%%%%%%%%%%%%%%%%%%%%%%%%%%%%
%%%%%%%%%%%%%%%%%%%%%%%%%%%%%%%%%%%%%%%%%%%%%%%%%%%%%%%%%%%%%%%%%%%%%%%%%%%%%%%%%%%%%%%%%%%%%%%%%%%%%%%%%%%%%%%%%%%%%%%%%%%%%%%%%%%%%%%%%%%%%%%%%%%%%%
%%%%%%%%%%%%%%%%%%%%%%%%%%%%%%%%%%%%%%%%%%%%%%%%%%%%%%%%%%%%%%%%%%%%%%%%%%%%%%%%%%%%%%%%%%%%%%%%%%%%%%%%%%%%%%%%%%%%%%%%%%%%%%%%%%%%%%%%%%%%%%%%%%%%%%
\section{Concluding remarks}
\label{sec4}

In this paper we have studied a VSR inspired modification of Maxwell-Chern-Simons electrodynamics.
The analysis consisted in formulating a SIM$(1)$--VSR topologically electrodynamics, 
with the expectation that the nonlocal effects would contribute not only as massive contributions
but rather in a significant way showing a distinct departure from the usual MCS theory.

We started with a brief construction of the SIM$(1)$ Abelian gauge symmetry.
Hence, with a proper definition for the wiggle field strength we proposed a SIM$(1)$--VSR MCS theory.
By adding an electrostatic source, we have determined the VSR-modified solution for the electric field.
In particular, we showed that at short distances, although the usual MCS contribution is still singular, as $r \rightarrow 0$,
the VSR-effects give a finite contribution in this case.

Next, the electrostatic field energy has been computed, and was used in order to compare the VSR contributions
in face of the usual MCS result.
At last, we have made use of the gauge-invariant formalism in order to compute the static potential
between opposite charges. Surprisingly, we found that VSR-effects contribute so that the usual (MCS) singular contribution
 for the potential can be suitably removed for a particular choice of the VSR parameter.
Hence, in addition to its regular form, the complete expression for the VSR modified (confining) potential shows
a prominent and health departure from the MCS theory as shown in Fig.\ref{fig2}.

%%%%%%%%%%%%%%%%%%%%%%%%%%%%%%%%%%%%%%%%%%%%%%%%%%%%%%%%%%%%%%%%%%%%%%%%%%%%%%%%%%%%%%%%%%%%%%%%

\subsection*{Acknowledgments}

R.B. thankfully acknowledges FAPESP for
support, Project No. 2011/20653-3.

\end{document}